\documentclass{article}

\usepackage{PRIMEarxiv}

\usepackage[utf8]{inputenc} 
\usepackage[T1]{fontenc}    
\usepackage{hyperref}       
\usepackage{url}            
\usepackage{booktabs}       
\usepackage{amsfonts}       
\usepackage{nicefrac}       
\usepackage{microtype}      
\usepackage{lipsum}
\usepackage{fancyhdr}       
\usepackage{graphicx}       
\graphicspath{{media/}}     

\title{Ethical Considerations and Policy Implications for Large Language Models: Guiding Responsible Development and Deployment

}

\author{
  Jianyi Zhang \\
  Affiliation \\
  Beijing Electronic Science and Technology Institute \\
  Beijing, China\\
  \texttt{zjy@besti.edu.cn} \\
  \And
  Xu Ji\\
  Affiliation \\
  Beijing Electronic Science and Technology Institute \\
  Beijing, China\\
  \texttt{jx4125@gmail.com} \\
   \And
  Zhangchi Zhao \\
  Affiliation \\
  Beijing Electronic Science and Technology Institute \\
  Beijing, China\\
  \texttt{1203657529@qq.com} \\
  \And
  Xiali Hei \\
  Affiliation \\
  University of Louisiana at Lafayette \\
  Lafayette, USA\\
  \texttt{xiali.hei@louisiana.edu} \\
  \And
  Kim-Kwang Raymond Choo \\
  Affiliation \\
  University of Texas at San Antonio \\
  Lafayette, USA\\
  \texttt{raymond.choo@utsa.edu} \\
}

\begin{document}
\maketitle

\begin{abstract}

This paper examines the ethical considerations and implications of large language models (LLMs) in generating content. It highlights the potential for both positive and negative uses of generative AI programs and explores the challenges in assigning responsibility for their outputs. The discussion emphasizes the need for proactive ethical frameworks and policy measures to guide the responsible development and deployment of LLMs.
\end{abstract}


The utility of large language models (LLMs), like ChatGPT, has been extensively demonstrated in a wide range of applications, including passing bar exams, writing full feature articles, and even generating complete website code. However, it is important to acknowledge the existence of risks and vulnerabilities associated with these models. Some instances have surfaced where ChatGPT was mis/ab-used to generate socially unacceptable content, such as promoting racial discrimination or facilitating misinformation and disinformation campaigns. To address this concern, OpenAI has implemented measures to mitigate the risks, such as installing a filter or classifier to identify dangerous prompts. Consequently, the output of ChatGPT has become more standardized, although human moderators are still necessary to identify and label inappropriate content~\cite{jo2023promise}.

Nonetheless, the flexibility of linguistic language can introduce complexities in the design of LLMs. One challenge arises from the fact that certain terminologies can possess different meanings or interpretations within different cultural or contextual contexts. In the following sections, we will provide a brief overview of these challenges.

\section*{System-role}\label{subsec2}

To adapt LLMs for a broader range of scenarios and enhance their usefulness in individualized tasks, some researchers have employed role-playing as an extension method, using LLMs like ChatGPT to generate a series of role-playing prompts~\cite{prompts}. Additionally, structured tests have been conducted to explore how different characters would respond to the same questions in a role-playing context~\cite{shen2023chatgpt}. However, it's important to note that these role-playing prompts can bypass the restrictions imposed by the classifier.

It is possible to make the model assume roles, placing it in specific scenarios where it generates data that would not be outputted otherwise. However, setting specific scenarios for ChatGPT increases the likelihood of encountering problematic content in the output. These issues may expose internal information of the LLM or go undetected, thus increasing the security risks associated with the language model. Methods like populating an Excel sheet can trick LLMs into generating training data-like content or act as a deceased grandmother who would read the Windows 10 Pro keys to fall asleep.

LLMs, including ChatGPT, offer users the ability to perform various tasks, such as multilingual translation. While OpenAI has implemented restrictions on input content to prevent violations of community rules~\cite{Moderation}, LLMs may still struggle with handling complex scenarios effectively. In translation tasks, for instance, if inappropriate content that violates legal, moral, or ethical standards is not adequately identified and restricted, it could be translated into multiple languages and widely disseminated, leading to significant consequences. Similarly, when requested to add negative or emotional elements to positive texts during translation, the emotional tone of the translated results may diverge from the intended content, potentially resulting in the generation of malicious or unintended output. OpenAI's official documentation acknowledges that "support for non-English languages is currently limited," indicating ongoing challenges in the realm of multilingualism that require continuous research and updates.

Different personas or characters within LLMs have diverse backgrounds, introducing significant flexibility to their responses. Consequently, the probability of encountering problems when these personas answer questions varies across different scenarios.

\section*{Perturbation}\label{subsec3}

Extensive research has been conducted on hints and prefixes associated with LLMs. Some studies have focused on maximizing the efficiency of language models, while others have delved into methods to bypass the toxicity filters of LLMs and generate potentially harmful content~\cite{si2022so}. These prefixes can assume various forms, including single words, phrases, sentences, or even paragraphs. Their purpose is to conceal crucial information that might trigger the language model's toxicity filter. Moreover, prefixes can adopt different grammatical structures and tenses, such as interrogative, declarative, or exclamatory sentences.

However, due to the rapid advancements in artificial intelligence, the prompts that trigger toxicity responses in language models may change in their expression, rendering current toxicity filters less effective in certain cases. Consequently, an attacker can search for an appropriate prefix that bypasses the filter and generates the desired content. Different prefixes can be incrementally employed, depending on the specific keywords for which the attacker intends to elicit toxic responses from the language model. By masking keywords using words or phrases—for example, using "objective" to conceal a controversial topic—an attacker can circumvent security measures. If more complex keyword phrases are required to exploit the security mechanism, the attacker may resort to using sentences or even paragraphs, which significantly increase the likelihood of bypassing the security measures. If the initial attempts are unsuccessful, the attacker can try employing longer paragraphs or multiple interactive sessions to mask content that policy guidelines prohibit. Given the language model's memory of chat history and interactive nature, the risk of obtaining toxic responses from the LLM grows as the conversation progresses and delves into deeper interactions. Additionally, due to the flexibility of language, different prompts provided to the LLM may result in output with similar toxic implications.

\section*{Image-related}\label{subsec3}

Image-related challenges arise in LLMs, primarily designed for text-based conversations, but capable of outputting images in Markdown format through API calls to image hosting websites, like ChatGPT. While this feature proves valuable for sharing visual information and enhancing the user experience, it carries certain risks. On one hand, it may inadvertently display inappropriate images related to pornography, gun violence, or drugs. On the other hand, when ChatGPT uses Markdown to display images without proper citation, it may lead to copyright disputes and harm the interests of creators.

While the ability to display images on the terminal is a valuable feature, it is crucial to exercise discernment and avoid displaying inappropriate images, as they can pose risks to social order and public opinion and may give rise to legal, moral, or ethical concerns. The release of multimodal models like GPT 4.0, which support image generation, necessitates increased attention to the ethical, moral, and safety issues arising from text-to-image functionality in the future. Baidu's LLM-based Chatbot ERNIE encountered ambiguities in the underlying code during text-to-image conversion. Testers, for example, used Chinese text to generate an image of a computer bus, but ERNIE generated an image of a transportation bus. This indicates a multilingual conversion error at the core of ERNIE after processing the text.

\section*{Hallucination}\label{subsec3}

In certain instances, LLMs may produce answers that sound plausible but are incorrect or nonsensical. This phenomenon is commonly referred to as "hallucination" in many articles~\cite{gusenbauer2023audit, vert2023will}. Our tests revealed that ChatGPT occasionally generates content that appears complete and objective, even without prior knowledge of the article's topic. However, it is important to acknowledge that AI can make mistakes, propagate misinformation, fabricate information, and generate realistic texts about events that never occurred~\cite{bbc2023}. Our investigations covered a wide range of domains, uncovering instances of misinformation and bias in ChatGPT's outputs. Some researchers have even identified errors in ChatGPT's determination of author contributions~\cite{van2023chatgpt}.

LLMs generate content that users may perceive as accurate and trustworthy information. However, if the generated content includes non-objective or unsupported information, it can be misleading and lead to incorrect judgments and decisions. If such content generated by LLMs is widely disseminated, it can significantly impact public opinion. Controversial topics may arise, and the inclusion of fictitious or misleading information can erode people's trust in AI, hindering the progress of the field. Additionally, it may affect people's usage and acceptance of LLMs, raising concerns about their objectivity.

Many individuals currently rely on the ChatGPT API to enhance their work efficiency. If the issue of hallucination is not effectively addressed, the existing problems and pitfalls may proliferate through the API interface. Therefore, it is essential to critically review and test the authenticity of LLMs' output content. Furthermore, adding citation sources for identified facts can help ensure that the generated content is trustworthy and valuable to human users.

\section*{Generation-related}\label{subsec3}

With the advancement of LLMs, the generated content produced by these models is being utilized in various domains, including marketing, education, cybersecurity, and more. It is crucial to ensure the ethical and responsible use of language models. Detecting whether the text is generated by an AI or a human is essential, as well as considering the societal impact of language model-generated texts.

With the ongoing evolution of LLM technology, the potential applications of language models are expanding. For instance, ChatGPT can serve as a versatile tool for marketing products and may eventually replace the need for professional marketers. Students may attempt to use ChatGPT for academic dishonesty, but the availability of high-performance detectors can help curb such incidents. Although ChatGPT-generated phishing emails have a lower success rate compared to human-written emails~\cite{hoxhunt2023}, this may change with the development of multimodal models like GPT 4.0. Attackers can leverage language models to generate fake news and rumors, disrupting established order and inducing panic among people. It is worth noting that ChatGPT can produce different outputs when given the same prompt, indicating inconsistent limitations that need improvement in terms of robustness~\cite{zhang2023one,wang2023robustness}.

LLMs serve as a tool for attackers, enhancing their abilities in developing malicious programs and increasing efficiency through rapid generation of mass production of malicious content. Although filters exist, attackers can generate a significant amount of small auxiliary codes using a language model and manually assemble them to create an attack system. Most current detectors only cover general text detection, and the development of specialized detectors targeting undesirable content, such as malicious code, rumors, or fake news, is still required.

\section*{Bias and Discrimination of Training Data}\label{sec4}

The functionality of the LLM relies on an extensive corpus of training data that encompasses various countries, cultures, and languages~\cite{seghier2023chatgpt}. This linguistic diversity equips the LLM with the competence to cater to the unique content needs of users across different languages. It excels in cross-language tasks, such as translation, as it can swiftly apply its capabilities in English to other language scenarios. However, the disparity between models trained in different languages extends beyond just the language itself. It encompasses factors like the country, culture, religion, politics, and ideological nuances associated with each language~\cite{seghier2023chatgpt}.

Therefore, when multiple language models are unified under a single model, they embody distinct aspects that reflect the diverse ideologies represented within the languages. Consequently, the responses to questions can diverge significantly depending on the underlying ideology. Bias and stereotypes arise from unfair judgments based on incorrect information and a lack of understanding toward individuals or subjects that differ from our own~\cite{bbcbard2023}.

For instance, people from different countries may hold disparate perspectives on a controversial issue, leading to variations in the training data available for these national languages. As a result, the LLM may exhibit distinct interpretations of a particular viewpoint, providing substantially different answers to the same question depending on the language in use. Attackers can exploit these differences to incite tensions between nations or groups with differing viewpoints. Additionally, the LLM's adaptability to a user's worldview highlights the non-consistent and non-deterministic nature of its underlying worldview.

\section*{Conclusion}\label{sec4}

The six categories discussed encompass the major security issues associated with LLMs, covering aspects related to input, output, and training data. Both the input and output components of LLMs are crucial to consider. Attackers leverage the input phase to prompt injection, exposing vulnerabilities in the generated content of LLMs. Consequently, careful review and evaluation of each generated component of LLMs become necessary.

While LLM classifiers exhibit high accuracy, the flexibility of language and diverse scenarios necessitate more than just prompt restrictions for effective protection. The temporary validity of prompt restrictions highlights the need for researchers to explore alternative approaches to enhance the classifier's protection system.

Aside from prompt injection, internal issues related to training data also contribute to the challenges. Relying solely on prompt filters is insufficient to achieve a perfect defense. Developing LLMs should prioritize addressing legal, moral, and ethical concerns. Legislation is required for AI products to strike a balance between fostering innovation and restraining unethical practices. Simultaneously, users must take responsibility and cultivate responsible usage habits. LLMs should diligently identify and reject any speech or description that harms others. This ensures the evolution of a scientifically sound research environment that upholds responsibility and promotes the well-being of society as a whole.


\bibliographystyle{unsrt}
\bibliography{main}

\end{document}